\documentclass[twoside]{article}
\usepackage{fleqn,espcrc2,epsfig}

\newcommand{\AmS}{{\protect\the\textfont2
  A\kern-.1667em\lower.5ex\hbox{M}\kern-.125emS}}

\title{Composite reweighting with Imaginary Chemical Potentials in SU(3)}

\author{ P. R. Crompton }

\begin{document}
\pagestyle{empty}
\setcounter{topnumber}{1}

\begin{abstract}
{We review the overlap pathology of the Glasgow reweighting method for finite density QCD, and discuss the sampling bias that effects the determination of the ensemble-averaged fugacity polynomial expansion coefficients that form the Grand Canonical Partition function. The expectation of the difference in free energies between canonical partition functions generated with different measures is presented as an indicator of a systematic quark number dependent biasing in the reweighting approach. The advantages of building up an unbiased polynomial expansion for the Grand Canonical Partition function through a series of parallel ensembles generated by reweighting with imaginary chemical potentials are then contrasted with those of addressing the overlap pathology through a secondary reweighting.}
\end{abstract}

\maketitle

\section{Introduction}
Interest in the numerical determination of BCS-type instabilities at the Fermi surface in QCD at high density has led to a recent resurgence of interest in the analytic continuation of Monte Carlo analyses and reweighting \cite{simon}\cite{olaf}.
For QCD the SU(3) lattice action \cite{kogut} becomes complex at nonzero chemical potential $\mu \neq 0$, and so the probabilistic importance sampling steps of a Markov chain are undefined prohibiting standard Monte Carlo techniques. The Glasgow method has given a useful means of quantifying the success of these approaches, where the $\mu$ dependence of the Grand Canonical Partition function $\Omega(\mu)$ is made semi-analytic through polynomial expansion in fugacity $z=e^{ \mu /T}$ \cite{ian}. 

Although the complex action issue is negated in generating a lattice ensemble in an accessible regime of the parameter space (eg. $\mu =0$ for SU(3)), the analysis is pathological. An unphysical onset transition in the quark number density $J_{o}$ at $\mu = \frac{1}{2}m_{\pi}$ is thought be a feature of quenched (or isolated configurations of) QCD at finite density, but in addition exponentially high statistics are required to generate physical results from dynamically generated ensembles. It is this second overlap pathology which also persists for QCD-like models with the Glasgow scheme in which the Goldstone modes of the theories are either degenerate or considerably heavier than $\frac{1}{2}m_{\pi}$ \cite{christine}\cite{me}\cite{simon2}, which hints further at a statistical origin. 

By evaluating the ratio of ensemble-averaged terms in these QCD-like models (for which the fermion determinant is positive definite) we were able to study the comparative effect of reweighting with the Glasgow method for ensembles generated at different values of chemical potential $\mu_{o}$, and in so doing develop a new reweighting approach for QCD.
\section{Glasgow Method}
The characteristic polynomial in fugacity is formed through the definition of the propagator matrix $P$ \cite{phil} from gauge links in the spatial $G$ and time directions $V$.
\begin{eqnarray}
P & = & \left( \begin{array}{cc} 
 -( G + 2im ) & 1  \\
-1	       & 0   \end{array}\right) V \\
{\rm{det}} M  & = & {\rm{det}}  ( G + 2im + 
V^{\dagger}e^{-\mu} + Ve^{\mu} ) \\
& = & e^{n_{c}n_{s}^{3}n_{t}\mu} \,\,{\rm{det}}  ( P - e^{-\mu} )\\
& = & e^{n_{c}n_{s}^{3}n_{t}\mu}\sum_{n=0}^{2n_{c}n_{s}^{3}n_{t}}
c_{n}e^{-n\mu} 
\end{eqnarray}
Since $V$ is an overall factor of $P$ the order of the expansion can be reduced by exploited the unitary symmetry $Z_{n_{t}}$ defined by $e^{2\pi ij / n_{t}}$ (where $j$ is integer). This relates the expansion terms to the GCPF $\Omega(\mu)$ through the Canonical Partition functions for $n$ quarks, represented by the suitably normalised ensemble-averaged expansion coefficients in a reweighted measurement. 
\begin{eqnarray}
\frac{\Omega(\mu)}{\Omega(\mu_{o})} & = & \sum_{n} \frac{\Omega_{n}}{\Omega(\mu_{o})} \, e^{n\mu/T} \\
\frac {\Omega_{n}} {\Omega(\mu_{o})} & = & \frac {\int DU {\displaystyle{\frac {c_{n}} { {\rm{det}} 
M(\mu_{o}) }}}
\,\, {\rm{det}} M(\mu_{o}) \,\, e^{-S_{g}}} {\int DU 
\,\, {\rm{det}} M(\mu_{o}) \,\, e^{-S_{g}}} \\
& = & \left\langle {\frac{{c_{n}}}{{\rm{det}} M(\mu_{o})}}
 \right\rangle_{\mu_{o}} \\
& = & \frac{\left\langle {\frac{c_{n}}{ {\rm{det}} M(\mu_{1}) } \frac{ {\rm{det}} M(\mu_{1}) }{ {\rm{det}} M(\mu_{o}) }}
\right\rangle_{\mu_{1}} }{ 
 \left\langle {\frac{ {\rm{det}} M(\mu_{1}) }{ {\rm{det}} M(\mu_{o}) }} \right\rangle_{\mu_{1}} } 
\end{eqnarray}

In addition to possible autocorrelations through the numerical evaluation of the ensemble there is also an inherent covariance between the numerator and denominator of a reweighted measurement which becomes exponentially dominant for significantly different ratios of weightings \cite{ferrenberg}. The Cauchy distribution of the ratio population formed by two independent gaussian distributions gives a simplistic picture of this behaviour. The joint density function integral for this ratio population diverges and the higher moments are thus undefined. Only for identical numerator and denominator populations does the joint density function then reduce to a gaussian, following suitable substitution. 

For SU(3) given the accessible region of the $T-\mu$ phase plane for importance sampling methods, the ratio of weightings given in Eqn.8 generally does differs from one when the expansion is reweighted to the vicinity of the critical line. Kurtosis in the measurement from the weighted average relative formed with the ratio of weightings is of particular concern for this scheme since the expansion coefficients vary over several hundred orders of magnitude. In SU(3)  for a $4^{4}$ volume typically at $\beta=5.04$,  $c_{n=0} \sim e^{0}$ and $c_{n=n_{c}n_{s}^{3}} \sim e^{440}$ and the kurtosis is thus uneven across the expansion. For the limited range of the suitably normalised expansion terms of order one (from an ensemble generated at some $\mu_{1}$), the numerator and denominator populations become equivalent and the averaging is then independent of the reweighting evaluation at $\mu_{o}$. Since these expansion coefficients are in general complex integrating over all gauge configurations cancellation occurs of all but the triality zero terms $( mod(j,n)$ $j=1,2 )$, but with a significant real-valued kurtosis this cancellation becomes inexact. The phase of the ratio of complex weightings is of less consequence given this $Z(3)$ tunneling and the equality of the transition probability for conjugate gauge configurations.

The related effect was noticed in the evaluation of $\Omega(i\nu)$ by reweighting using dynamical ensembles generated at various values of $i\nu_{o}$ for the Hubbard model \cite{mark}. To address the kurtosis with complex weightings for QCD we could envisage recovering each $\Omega_{N}$ through the Fourier transform of $\Omega (i\nu)$, using a reweighted fugacity expansion with an ensemble generated at some $i\nu_{o}$. With this scheme for $\Omega(i\nu)$ to be reliably sampled through reweighting we would have to evaluate ensembles for a series of values of $i\nu_{o}$, but the only non-vanishing term of the fugacity expansion from the Fourier transform is $c_{N}$, and the identification of a unique value of $\nu_{o}$ for which the sampling of $c_{n}$ is effective is therefore explicit. 

\begin{equation}
\frac{\Omega_{N}}{\Omega(i\nu_{o})} = \frac{1}{2\pi} \int \! d\nu \,\sum_{n} \left\langle {\frac{{c_{n}}}{{\rm{det}} M(i\nu_{o})}}
 \right\rangle_{i\nu_{o}} e^{\nu (n-N)}
\end{equation}
\begin{equation}
{\rm{det}}M(\mu_{1}) = c_{n} = \prod_{n=1}^{2n_{c}n_{s}^{3}} ( z - \lambda_{n} ) 
\end{equation} 

However, rather than evaluate a costly series of ensembles generated for a range of values of $i\nu_{o}$ to ensure the analyticity of the fugacity polynomial we can make explicit measurements for the numerator and denominator to address the kurtosis of complex weightings. The eigenvalue problem to identify $\mu_{1}$ for a given $c_{n}$ is first solved, and then the numerator evaluated with these normalisations for each term of the expansion. The product of the weighting ratios for each consecutive value of $\mu_{1}$ is then evaluated, thus ensuring the denominator evaluation is also of order one. 

Our results for 10,000 configurations of a $4^4$ lattice with four staggered quark flavours at $\beta=5.04$ and $m=0.1$ are presented in Fig.1, with the Lee-Yang zeros of the GCPF evaluated in the complex-$\mu$ plane. Our improvement scheme reveals a rich finite density phase structure, with two clear transition points (where the zeros approach the real axis). The first is in agreement with \cite{fodor+katz} ( Re($\beta_{o}$)=4.98(9) ), and the second also identified in \cite{never} appears from our preliminary scaling analysis to be associated with lattice saturation. Notably, the critical feature at $\mu\sim0.4$ appears to correspond to $\frac{1}{2}m_{\pi}$ for this bare mass \cite{ian2}, and it will be interesting to repeat the analysis at smaller values of $m$.

\begin{figure}
\centering{\epsfig{file=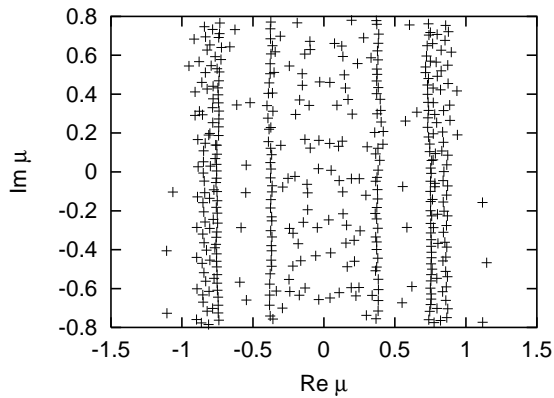,height=5.4cm, width=7.7cm}}
\vspace{-1.4cm}
\caption{Lee-Yang zeros for QCD evaluated in the complex-$\mu$ plane for a 
$4^{4}$ lattice at $\beta=5.04$ .}
\end{figure}
\section{Summary}
Each analytic continuation scheme reliant on reweighting can be pictured as a Taylor expansion about the zero density ensemble, and through the Glasgow scheme we can identify the range of validity of ensemble-averaging for complex weightings. We have addressed the inherent kurtosis of measurement by redefining our ensemble-averaging and can now more reliably evaluate the Grand Canonical Partition function, ensuring analyticity in the fugacity variable. Using in addition reweighting simultaneously in $\beta$ we could also now reliably reweight to the entire $T-\mu$ plane (up to lattice saturation for finite volumes) given the independence on Monte Carlo measure we have established.

The Lee-Yang zeros we have evaluated from our new scheme appear to approach the real axis in manner accordant with the finite volume scaling circle theorem of \cite{Lee-Yang}. However, it remains to be seen if we can make useful comments on diquark condensation at finite density, and evaluate the critical exponents of this transition following approaches suggested in \cite{vafa1}\cite{vafa2} through the finite volume scaling of the edge singularity revealed with this analysis.

\end{document}